\documentclass{jfm}
\usepackage{graphicx}
\usepackage{epstopdf, epsfig}
\usepackage{amsmath}

\usepackage{xcolor}
\usepackage{bm}
\definecolor{b1}{rgb}{0.743,0.766,0.918}
\definecolor{b2}{rgb}{0.487,0.533,0.836}
\definecolor{b3}{rgb}{0.230,0.299,0.754}
\definecolor{r1}{rgb}{0.941,0.803,0.830}
\definecolor{r2}{rgb}{0.882,0.606,0.660}
\definecolor{r3}{rgb}{0.824,0.407,0.490}
\definecolor{r4}{rgb}{0.765,0.213,0.320}
\definecolor{r5}{rgb}{0.706,0.016,0.150}
\definecolor{p1}{rgb}{0.813,0.854,0.854}
\definecolor{p2}{rgb}{0.625,0.708,0.708}
\definecolor{p3}{rgb}{0.438,0.563,0.563}
\definecolor{p4}{rgb}{0.292,0.354,0.417}
\definecolor{p5}{rgb}{0.146,0.146,0.271}

\shorttitle{Spectrum and variance of wall-normal velocities}
\shortauthor{M. Heisel, R. Deshpande, and G. G. Katul}

\title{On the wall-normal velocity variance in canonical wall-bounded turbulence}

\author{Michael Heisel\aff{1}
  \corresp{\email{michael.heisel@sydney.edu.au}},
  Rahul Deshpande\aff{2},
 \and Gabriel G. Katul\aff{3}}

\affiliation{\aff{1}School of Civil Engineering, University of Sydney, Sydney, NSW 2006, Australia
\aff{2}Department of Mechanical Engineering, University of Melbourne, Parkville, VIC 3010, Australia
\aff{3}Department of Civil and Environmental Engineering, Duke University, Durham, NC 27708, USA}

\begin{document}

\maketitle

\begin{abstract}

The variance and spectra of wall-normal velocities are investigated for direct numerical simulations of turbulent flow in a channel, pipe, and zero-pressure-gradient boundary layer across a decade of friction Reynolds numbers. Spectra along the spanwise wavenumber have a pronounced peak well described by the turbulent dissipation rate and the local shear stress throughout the bottom half of the boundary layer. Deviations in the local stress from the surface shear velocity $U_\tau$ account for almost all of the differences in wall-normal velocity variance observed across different canonical flows, including for plane Couette flow. The dependence on the local stress is attributed to the fact that wall-normal motions are predominately `active' per Townsend's attached eddy hypothesis and directly contribute to the local shear stress, noting this hypothesis assumes simplified ideal conditions with constant turbulent shear stress. A semi-empirical fit applied to the Reynolds number dependence of the variance matches the simulations across the lower half of the boundary layer and aligns with observed values in the literature. The fit extrapolates to a value between 1.45 and 1.65 times the local shear stress in the high-Reynolds-number limit, consistent with previous predictions relative to $U_\tau$ including for the vertical velocity in the near-neutral atmospheric boundary layer. However, universality in the exact proportional constant is precluded by small discrepancies in the variances corresponding to dissimilarity in the low-wavenumber contributions across different flow configurations and wall-normal positions. We speculate the dissimilarity is due to relatively weak ‘inactive’ wall-normal motions that are excluded from Townsend’s original hypothesis.

\end{abstract}

\section{Introduction}
\label{sec:intro}

The Attached Eddy Hypothesis (AEH) of A. A. \citet{Townsend1976}, along with several decades of subsequent investigations \citep{Marusic2019}, is a key advancement in describing turbulent statistics in canonical wall-bounded flows. The AEH provides a phenomenological link between the turbulence, the logarithmic mean velocity profile, and the logarithmic decay in the streamwise and spanwise velocity variances. These trends have all been supported by measurements at sufficiently high Reynolds numbers across various canonical wall-bounded flows \citep[see, e.g.,][]{Jimenez2008,Hultmark2012,Marusic2013,Orlandi2015,Lee2015} including the near-neutral atmospheric surface layer \citep{Puccioni2023,Qin2025}.

The population of attached eddies envisioned by Townsend also produces a constant wall-normal velocity variance $\overline{w^{\prime 2}}$ within the logarithmic region of the boundary layer:

\begin{equation}
\dfrac{\overline{w^{\prime 2}}(z)}{ U_{\tau}^2} = B_3,
\label{Eq1_1}
\end{equation}

\noindent which is proportional to the surface shear (friction) velocity $U_\tau = \sqrt{\tau_w/\rho}$ corresponding to the wall shear stress $\tau_w$ and fluid density $\rho$. Wall-normal profiles of $\overline{w^{\prime 2}}$ tend to vary only weakly within the logarithmic region \citep{Morrill2015,Qin2025} and there is evidence for a constant third-order skewness \citep{Buono2024a}. Yet, there is no conclusive universal value for $B_3$ from observations and Eq. \eqref{Eq1_1} has the least experimental support among the statistics predicted by the AEH \citep{Buono2024b}.

\begin{figure}
\centerline{\includegraphics{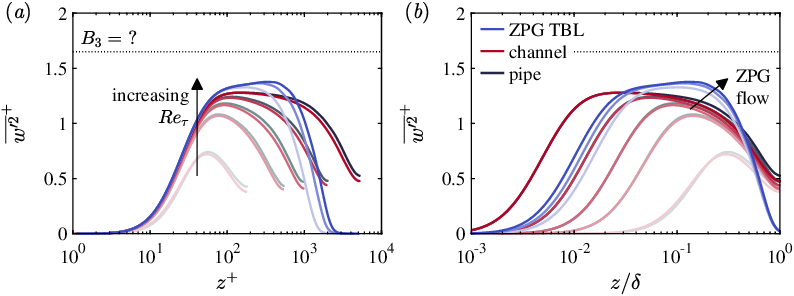}}
\caption{Profiles of the wall-normal velocity variance $\overline{{w^\prime}^2}$ normalized by the surface shear velocity $U_{\tau}$: (\textit{a}) wall-normal position $z$ in viscous units $z^+=z U_\tau / \nu$; (\textit{b}) $z$ in outer units relative to the boundary layer thickness $\delta$. In all figures, the data are from direct numerical simulations (DNS) \citep{Sillero2013,Lee2015,Yao2023}, with color indicating the flow case and shade corresponding to the friction Reynolds numbers $Re_\tau = \delta U_\tau /\nu$ given in Table \ref{Tbl}.}
\label{Fig1}
\end{figure}

There are two primary challenges in determining the value of $B_3$. The first is that the AEH models an asymptotically high-Reynolds-number wall-bounded flow with negligible viscosity effects within the logarithmic region. Observations of $\overline{w^{\prime 2}}$ in available simulations and laboratory-scale experiments are limited to finite Reynolds number and must account for its dependence. An example is shown in Fig. \ref{Fig1}(\textit{a}) for direct numerical simulations (DNS) of flows in a zero-pressure-gradient (ZPG) flat plate boundary layer \citep{Sillero2013}, channel \citep{Lee2015}, and pipe \citep{Yao2023}. The $\overline{w^{\prime 2}}$ for channel and pipe flows is nearly constant across intermediate distances, but the magnitude increases with the friction Reynolds number $Re_\tau = U_\tau \delta / \nu$ with $\delta$ being the boundary layer thickness and $\nu$ the kinematic viscosity.

Perry and co-authors \citeyearpar{Perry1986,Perry1990} estimated this Reynolds-number dependence by considering the finite scale separation between the wall parameters and the viscous motions in the spectrum of $w$, resulting in a correction for $B_3$ as a function of $z^+ = z U_\tau / \nu$. An alternative approach was recently proposed by \citet{Spalart2021}, who used extrapolation of profiles at different Reynolds numbers following the technique of \citet{Luchini2017}. The method accurately describes channel data \citep{Hoyas2022}, but is empirical in nature and is not linked to the flow phenomenology. These previous efforts collectively have not produced a conclusive Reynolds number correction to Eq. \eqref{Eq1_1} and further research on the matter is warranted.

The exclusion of viscous effects in the AEH applies also to the Reynolds shear stress $\overline{u^\prime w^\prime}$. The assumption of a constant turbulent stress with negligible viscous stress contribution is an ansatz of the AEH:

\begin{equation}
\dfrac{-\overline{u^\prime w^\prime }(z)}{ U_{\tau}^2} = 1.
\label{Eq1_2}
\end{equation}
    
\noindent Equation \eqref{Eq1_2} further neglects any decrease in the total stress profile below $U_\tau^2$ as the wall-normal distance increases. Given that $\overline{w^{\prime 2}}$ and $\overline{u^\prime w^\prime}$ have both been postulated to receive contributions solely from the `active' motions per  \citeauthor{Townsend1976}'s (1976) AEH, viscous stresses and a decreasing total stress can directly impact $\overline{w^{\prime 2}}$. This has been demonstrated using simulations of channel flows under different Reynolds numbers and forcing conditions, where variability in $\overline{w^{\prime 2}}$ throughout most of the boundary layer follows $\overline{u^\prime w^\prime}(z)$ more closely than $U_\tau$ \citep{Tuerke2013,Lozano2019}. However, the effect of the stress profile has not been closely investigated in the context of $B_3$ and Eq. \eqref{Eq1_1}.

The second challenge to determining $B_3$ is an observed difference in the variance between wall-bounded flow configurations. As seen in Fig. \ref{Fig1}, the ZPG boundary layer has higher $\overline{w^{\prime 2}}$ relative to $U_\tau$ and a peak that occurs farther from the surface. This difference is often discussed in terms of enclosed versus non-enclosed flows. \citet{Jimenez2008} conducted a survey of available datasets and acknowledged the different amplitudes (see, e.g., their Fig. 4). A similar survey by \citet{Buschmann2009} discussed the different wall-normal position of the variance peak, but noted limited differences in the amplitudes. An explanation for these trends has not been offered in these previous studies, however, and it remains an open question regarding the wall-normal variance.

The combination of the Reynolds-number dependence and the different profile for ZPG boundary layers has led to a range of reported $B_3$ values for the high-$Re$ limit. \citet{Spalart1988} found $B_3 \approx$ 1.75 empirically from early DNS at low Reynolds numbers. \citet{Perry1990} refined their spectral approach with a model spectrum \citep{Kovasznay1948} and fitted $B_3\approx 1.6$ to experimental ZPG boundary layer data. \citet{Hafez1991} revised this value to $B_3\approx 1.78$ using further measurements, and the higher $B_3$ agrees well with measurements in high-$Re$ laboratory facilities \citep{Morrill2015} as well as the atmospheric surface layer \citep{Kunkel2006}. Empirical fits to a wider range of atmospheric measurements suggest $B_3\approx$ 1.55 to 1.7 in neutral conditions \citep{Panofsky1977,Garratt1992}. Recent experimental pipe flow data are also consistent with a higher $B_3$ value close to 1.85 \citep{Orlu2017}.

The range of reported values $B_3\approx$ 1.5 to 1.85 is promising in terms of being relatively narrow. However, reaching a consensus on a more precise value will require addressing the open questions and challenges in the dependencies of the wall-normal variance. Accordingly, the goal of this work is to further investigate the three questions indicated in Fig. \ref{Fig1}: (1) what is the correct formulation for the Reynolds number dependency? (2) why do ZPG boundary layers have a higher variance relative to $U_\tau$? (3) what is the value of $B_3$ in the high-$Re$ limit and is it universal?

The same DNS featured in the Fig. \ref{Fig1} example are used throughout the study. The focus of the analysis is to identify the dependencies of the spectrum for the wall-normal velocity. The integrated spectrum is then used to assess the implications of the observed dependencies on the variance statistics, including for the questions given above. The remainder of the article is outlined as follows: the methodology is detailed in \S \ref{sec:method}, including an overview of the simulations and spectrum analysis; results are then presented in \S \ref{sec:results}; the implications for the motivating questions are discussed in \S \ref{sec:discuss}; concluding remarks are given in \S \ref{sec:summary}.

\section{Methodology}
\label{sec:method}

\subsection{Simulations}

Three published DNS datasets are used here: ZPG flat plate boundary layer \citep{Sillero2013}, channel flow \citep{Lee2015}, and pipe flow \citep{Yao2023}. Each dataset includes a range of friction Reynolds numbers that are listed in Table \ref{Tbl} and shown as different color shades in Fig. \ref{Fig1}. All statistics and variables directly follow the published datasets, except where noted below.

\begin{table}
\begin{center}
\def~{\hphantom{0}}
\begin{tabular}{rcll}
Dataset & Symbol                            & $Re_\tau$ & Source               \\
\hline
        & \textcolor{b2}{$\textbf{---}$}  & 1\,310    &                      \\
ZPG TBL	& \textcolor{b2}{$\textbf{---}$}  & 1\,710	& \citet{Sillero2013}  \\
        & \textcolor{b3}{$\textbf{---}$}  & 1\,990    &                      \\
\hline
        & \textcolor{r1}{$\textbf{---}$}  & 180       &                      \\
        & \textcolor{r2}{$\textbf{---}$}  & 550   	&                      \\
channel & \textcolor{r3}{$\textbf{---}$}  & 1\,000    & \citet{Lee2015}      \\
        & \textcolor{r4}{$\textbf{---}$}  & 2\,000    &                      \\
        & \textcolor{r5}{$\textbf{---}$}  & 5\,200   	&                      \\
\hline
        & \textcolor{p1}{$\textbf{---}$}  & 180       &                      \\
        & \textcolor{p2}{$\textbf{---}$}  & 550   	&                      \\
pipe    & \textcolor{p3}{$\textbf{---}$}  & 1\,000    & \citet{Yao2023}      \\
        & \textcolor{p4}{$\textbf{---}$}  & 2\,000    &                      \\
        & \textcolor{p5}{$\textbf{---}$}  & 5\,200   	&                      \\
\end{tabular}
\caption{Previously published direct numerical simulation (DNS) datasets of zero-pressure-gradient turbulent boundary layers (ZPG TBL), channels, and pipe flows used in the present analysis.}
\label{Tbl}
\end{center}
\end{table}

For the ZPG flat plate dataset, three intermediate runs between $Re_\tau$ = 1310 and 1990 are excluded here for clarity in later figures, and on account of the relatively modest $Re_\tau$ difference across all runs compared to the other cases. The reported boundary layer thickness $\delta_{99}$ based on the convergence of the mean flow to free-stream conditions is not used. Rather, $\delta$ is determined as the $z$ position where the local total shear stress is $\tau(z) = 0.0015\tau_w$. Defining $\delta$ based on the vanishing stress is more appropriate for comparison between flow cases due to the relevance of the stress profile to the later analysis; for the  enclosed cases, the stress is zero at the channel half-height and pipe center that correspond to $\delta$.
The present TBL thickness definition yields $\delta$ values within 1\% of estimates from the phenomenological definition given recently by \citet{Lozier2025}, which is nominally $\gtrsim$\:25\% than the value of $\delta_{99}$ \citep{Morrill2015,Lozier2025}. 

Datasets with higher Reynolds numbers have been presented in the literature for some of the enclosed flow configurations \citep[see, e.g.,][]{Pirozzoli2021,Hoyas2022}. The datasets in Table \ref{Tbl} are used instead due to availability of spanwise wavenumber spectra at the time of the analysis.

\subsection{Spanwise wavenumber spectra}

One-dimensional velocity spectra are traditionally studied as a function of the streamwise wavenumber $k_x$. Along this direction -- and also for time series -- the wall-normal velocity spectrum $E_{ww}(k_x)$ is approximately constant at the lowest wavenumbers before undergoing an extended transition to the inertial subrange of scales at higher $k_x$  \citep{Ayet2020}. The number of contributing factors to the shape of this transition make it infeasible to identify the appropriate length scale for this region of the spectrum.

\begin{figure}
\centerline{\includegraphics{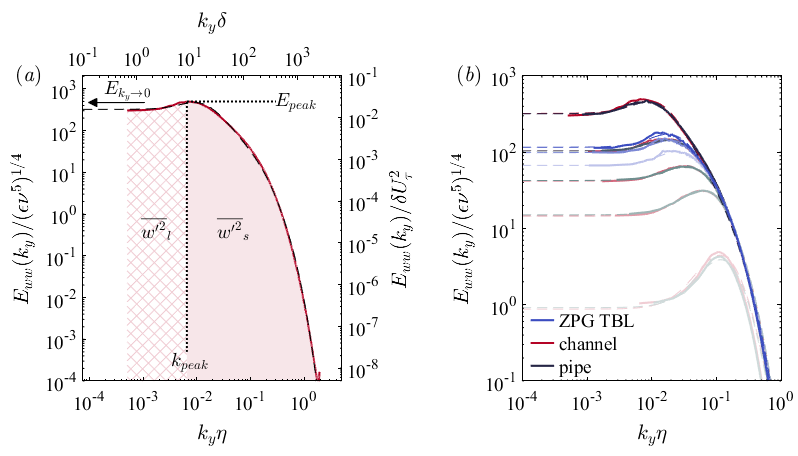}}
\caption{Example wall-normal velocity spectra $E_{ww}$ as a function of spanwise wavenumber $k_y$ for $z/\delta =$ 0.1, where the dashed lines are fits using the model von K{\'a}rm{\'a}n spectrum in Eq. \eqref{Eq2_1}. (\textit{a}) for the $Re_\tau=$ 5\,200 channel flow case \citep{Lee2015} to demonstrate the wavenumber $k_{peak}$ and energy density $E_{peak}$ of the spectrum peak, the low-wavenumber plateau $E_{k_y \to 0}$, and the contribution of large-scale motions $\overline{{w^\prime}^2}_{l} = \int_0^{k_{peak}} E_{ww} dk_y$ and small-scale motions $\overline{{w^\prime}^2}_{s} = \int_{k_{peak}}^\infty E_{ww} dk_y$ to the wall-normal variance. (\textit{b}) for all cases in Table \ref{Tbl} to demonstrate the model spectrum fit.}
\label{Fig2}
\end{figure}

In contrast, the spectrum $E_{ww}(k_y)$ as a function of the spanwise wavenumber $k_y$ exhibits a distinct peak at intermediate scales as seen in Fig. \ref{Fig2}(\textit{a}). There is a clearer segregation of motions in the spanwise spectrum, where the peak and low-wavenumber limit have different behaviors and can be individually assessed. Accordingly, $E_{ww}(k_y)$ statistics are investigated in this study. The later analysis was repeated using the peak in the premultiplied spectrum $k_y E_{ww}$ which is commonly employed in the literature for streamwise wavenumber spectra that otherwise lack a peak \citep[see, e.g.,][]{Baidya2017}. The same spectral scaling is observed for both approaches, but the uncompensated spectrum is less affected by the viscous cutoff and is the more robust option for this study. The spectra for the ZPG flat plate case are available at a limited number of $z$ positions, such that wall-normal profiles of spectral properties are coarse for this case in later figures.

Unlike a channel and flat plate that have a fixed length along $y$, a pipe geometry has a decreasing cross-flow distance (i.e., circumference) with increasing wall-normal distance. To account for this difference, the tangential wavenumber $k_t$ for the pipe flow must be converted to spanwise wavenumber as $k_y = k_t/(1-z/R)$ where $R=\delta$ is the pipe radius.

The $E_{ww}(k_y)$ spectra are well described by a model von K{\'a}rm{\'a}n spectrum \citep{Diederich1957}:

\begin{equation}
E_{ww}(k_y) = a \frac{1 + b(c k_y)^2}{\left[ 1+ (c k_y)^2 \right]^{11/6} } f(k_y\eta),
\label{Eq2_1}
\end{equation}

\noindent where $a$, $b$, and $c$ are fitted. The dissipative cutoff occurring at the Kolmogorov microscale $\eta$ is $f(k_y\eta)= \exp ( -\beta [ \{ (k_y \eta)^4 +0.15^4 \} ^{1/4}-0.15 ] )$ \citep{Pope2000} with $\beta=$ 5.2 following experimental evidence \citep{Saddoughi1994}, and 0.15 chosen empirically based on the alignment with the DNS spectra seen in Fig. \ref{Fig2}(\textit{b}).

Equation \eqref{Eq2_1} is a general form of the von K{\'a}rm{\'a}n spectrum because the integral length $L$ and root-mean-square velocity are not imposed. The scaling properties of the spectrum are a goal of the analysis and no parameters are assumed aside from $\eta$. The primary use of the model in Eq. \eqref{Eq2_1} is to estimate the low-wavenumber limit $E_{k_y \to 0}=a$ that corresponds to the energy splashing region captured within the DNS domain \citep{Ayet2020,Katul2013}. It is noted that the wavenumber $k_{peak}$ associated with the spectral peak may be determined analytically when assuming $f(k_{peak}\eta)=1$ (i.e. high Reynolds number flow). This wavenumber location is determined by setting $dE_{ww}(k_y)/dk_y=0$ to yield

\begin{equation}
k_{peak} = \frac{\sqrt{6b-11}}{c\,\sqrt{5 b}}.
\label{Eq2_2}
\end{equation}

This finding implies that (i) $b>11/6$, which is satisfied by $b=8/3$ in the original spectrum model \citep{Diederich1957}; and (ii) $k_{peak}$ is insensitive to $a$. While the leading factor does not determine the wavenumber of the peak, $E_{ww}(k_y=0) = a = 2 \overline{{w^\prime}^2} L_w /\pi$ relates to both the variance and the integral length scale $L_w$ of $w$ defined from the autocorrelation function. This definition for $E_{ww}(k_y=0)$ and $a$ results directly from the spectrum's relation to the autocorrelation, which corresponds to $L_w$ for $k_y=0$.

\subsection{Flow parameters}
\label{subsec:params}

A direct outcome of the attached eddy population in the AEH is that $z$ and $U_\tau$ are the appropriate scales for the energy-containing production region in the spectrum of turbulence. These scales are accordingly used in previous studies on the spectrum of $w$ \citep{Perry1986,Perry1990}, and are included in this analysis as a basis for comparison.

The scales $z$ and $U_\tau$ are strictly applicable to the logarithmic region in the high-Reynolds-number limit, and the parameters must be generalized outside of these conditions. To this end, Davidson and coauthors \citeyearpar{Davidson2006,Davidson2014} proposed a length $U_\tau^3/\epsilon$ based on the local dissipation rate $\epsilon$. This length provides a mathematically consistent transition into the inertial subrange of the spectrum where $\epsilon$ is the sole scaling parameter, and accounts for the fact that lengths besides $z$ may contribute to setting $\epsilon$ depending on the conditions \citep{Davidson2014}. In practice, $U_\tau^3/\epsilon$ asymptotes to $\kappa z$ in the high-$Re$ logarithmic region \citep{deSilva2015}, where $\kappa$ is the von K{\'a}rm{\'a}n constant, and in these conditions the premultiplied peak in the $w$ spectrum is well described by $z$ \citep{Kunkel2006,Baidya2017}. However, $U_\tau^3/\epsilon$ outperforms $z$ in delineating the start of the inertial sublayer for lower Reynolds numbers \citep{Davidson2006,Heisel2022}, within the roughness sublayer \citep{Davidson2014,Ghannam2018}, and for stratified flow \citep{Chamecki2017}.

The velocity scale can also be generalized by considering a local-in-$z$ shear velocity corresponding to the decaying stress profile \citep{Tuerke2013,Lozano2019}. This is a relaxation of Eq. \eqref{Eq1_2} in the AEH to account for conditions when the local stress decreases below the surface value given by $U_\tau$. The generalization allows $\overline{w^{\prime 2}}$ to maintain its relation with the local stress as discussed in \S \ref{sec:intro}. In atmospheric flows, this approach is known as ``local scaling'' and it provides improved mean velocity similarity under stratified conditions \citep{Nieuwstadt1984,Holtslag1986,Heisel2023}. The local shear velocity is used here also to define the dissipation-based length scale discussed above, leading to the following revised scaling parameters:

\begin{align}
    u_{\tau z}^2 &= -\overline{u^\prime w^\prime } + \nu \frac{dU}{dz}, \label{Eq2_3}\\
    \ell_\epsilon &= \frac{u_{\tau z}^3}{\epsilon}. \label{Eq2_4}\
\end{align}

\noindent Here, the local shear velocity $u_{\tau z}$ is defined from the total shear stress $\tau(z)$ comprising a turbulent and viscous component. Including the viscous component leads to improved agreement with the spectra compared to using only the turbulent Reynolds shear stress.

\begin{figure}
\centerline{\includegraphics{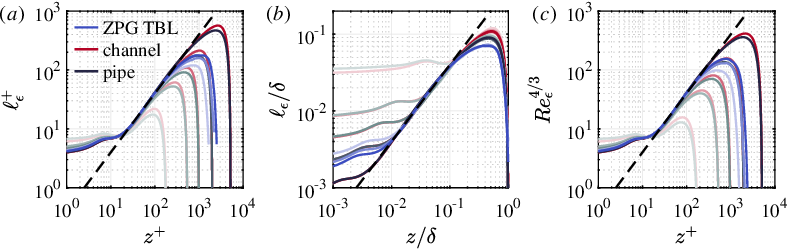}}
\caption{Profiles of the dissipation-based length $\ell_\epsilon$ defined in Eq. \eqref{Eq2_4}: (\textit{a}) $\ell_\epsilon$ in viscous units; (\textit{b}) $\ell_\epsilon$ in outer units; (\textit{c}) the Reynolds number $Re_\epsilon = \ell_\epsilon / \eta$ representing separation between length scales. The dashed lines in each panel all correspond to $\kappa z$.}
\label{Fig3}
\end{figure}

The relation between $z$ and $\ell_\epsilon$ from Eq. \eqref{Eq2_4} is shown in Fig. \ref{Fig3}. The DNS align closely with $\ell_\epsilon = \kappa z$ from the top of the buffer layer near $z^+ \approx$ 30 in Fig. \ref{Fig3}(\textit{a}) to the top of the logarithmic region at $z/\delta \approx$ 0.1 in Fig. \ref{Fig3}(\textit{b}). The only exception is for $Re_\tau=$ 180, where there is insufficient separation between the buffer and wake regions for a logarithmic region to develop. The peak of $\ell_\epsilon$ near $z/\delta \approx$ 0.5 in Fig. \ref{Fig3}(\textit{b}) is different between the DNS cases owing to well-known differences in outer layer structure, along with the presence (absence) of turbulent/non-turbulent interfaces in non-enclosed (enclosed) flows \citep{Monty2009}. 

An important consideration to the later analysis is scale separation between the production and dissipative regions in the spectrum of turbulence. If $\ell_\epsilon$ is the appropriate parameter for delineating the production region, this separation can be directly represented by the dissipation-based Reynolds number $Re_\epsilon = \ell_\epsilon / \eta$, where $\eta=(\nu^3/\epsilon)^{1/4}$ is the Kolmogorov length scale. From the definitions of $\eta$ and $\ell_\epsilon$ in Eq. \eqref{Eq2_4}, the Reynolds number can be rewritten as $Re_\epsilon = (\ell_\epsilon u_{\tau z} / \nu)^{3/4}$. As seen in Fig. \ref{Fig3}(\textit{c}), the value can be approximated as $Re_\epsilon \approx ( \kappa z^+)^{3/4}$ in the logarithmic region where $\ell_\epsilon \approx \kappa z$ and $u_{\tau z} \sim U_\tau$.  These findings are consistent with studies that connect the degrees of freedom of a (3-D) turbulence cascade to $[(L_i/\eta)^{3/4}]^3$ \citep{Constantin1985,Landau2013}, when setting the integral scale $L_i$ to $\ell_\epsilon$.

\section{Results}
\label{sec:results}

\subsection{Spanwise spectrum properties}

Example spectra $E_{ww}(k_y)$ are compared at fixed positions of $z$ in viscous and outer units to evaluate whether the parameters introduced in \S \ref{subsec:params} can account for the spectrum properties. Then, trends in the peak and low-wavenumber spectrum properties are evaluated across the full boundary layer thickness for all DNS cases. These properties are introduced and visualized in Figure \ref{Fig2}(\textit{a}).

\begin{figure}
\centerline{\includegraphics{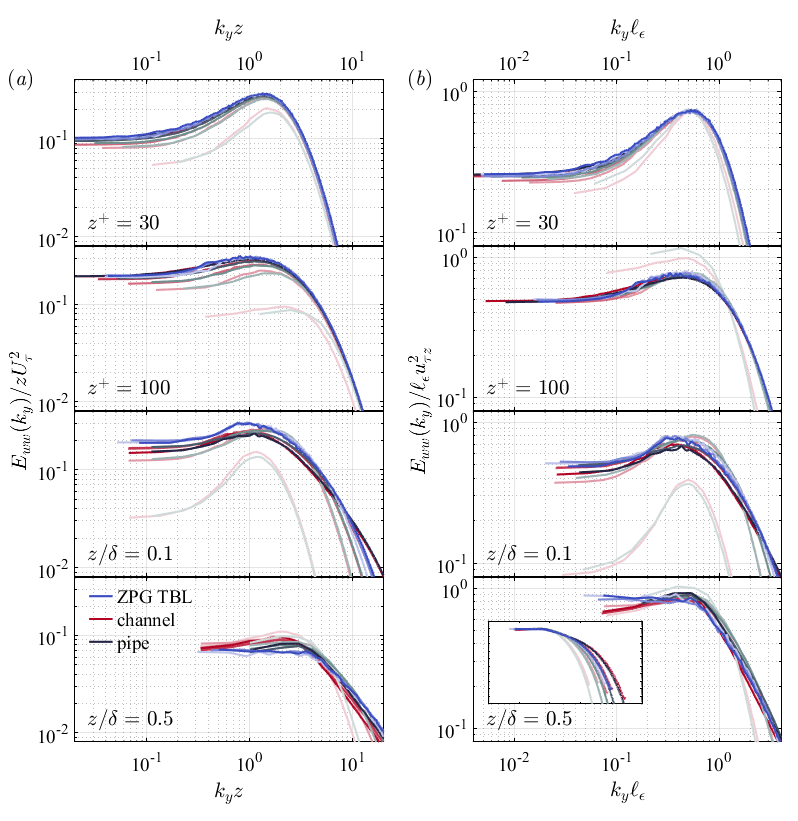}}
\caption{Comparison of $E_{ww}(k_y)$ at fixed $z$ positions in viscous and outer units: (\textit{a}) using traditional wall-scaling parameters $U_{\tau}^2$ and $z$; (\textit{b}) using local-in-$z$ parameters with shear velocity $u_{\tau z}^2 = -\overline{u^\prime w^\prime} + \nu \partial U / \partial z$ and $\ell_\epsilon$. The inset plot in (\textit{b}) shows the full spectrum with dissipative scales for reference.}
\label{Fig4}
\end{figure}

Figure \ref{Fig4} compares the DNS $E_{ww}(k_y)$ spectra at four positions: $z^+=$ 30, $z^+=$ 100, $z/\delta=$ 0.1, and $z/\delta=$ 0.5. The spectra are normalized by the wall-scaling parameters $U_\tau$ and $z$ in Fig. \ref{Fig4}(\textit{a}) and by the general parameters $u_{\tau z}$ and $\ell_\epsilon$ in Fig. \ref{Fig4}(\textit{b}). Normalization with the general parameters yields a better collapse of the spectrum peak at each of the wall-normal positions shown, indicating that $u_{\tau z}$ and $\ell_\epsilon$ more accurately define the wavenumber $k_{peak}$ and energy density $E_{peak}$ of the peak across a broad range of Reynolds numbers.

At $z^+=$ 30, $z^+=$ 100, and $z/\delta=$ 0.1, the spectral peaks of the highest $Re_\tau$ cases are well described by both wall scaling in Fig. \ref{Fig4}(\textit{a}) and the general parameters in Fig. \ref{Fig4}(\textit{b}). The alignment of both options under high-$Re$ conditions is consistent with the close relation $\ell_\epsilon = \kappa z$ seen in Fig. \ref{Fig3} and previous experimental observations that wall scaling is applicable under these conditions \citep{Kunkel2006,Baidya2017}. The discrepancies at lower $Re_\tau$ in Fig. \ref{Fig4}(\textit{a}) are accounted for using the more general length and velocity parameters, indicating that $u_{\tau z}$ and $\ell_\epsilon$ can account for finite-Reynolds-number effects in the spectral peak. Further, the general parameters lead to reduced differences in the spectrum amplitudes of the ZPG and enclosed flows at $z/\delta=$ 0.1. The only deviation in Fig. \ref{Fig4}(\textit{b}) occurs for the $Re_\tau=$ 180 cases at $z^+=$ 100 and $z/\delta=$ 0.1. At these positions for low Reynolds numbers, both viscous and $\delta$-scale effects are substantial.

At $z/\delta=$ 0.5, the spectrum peak remains well defined by $\ell_\epsilon$ as seen in Fig. \ref{Fig4}(\textit{b}). In contrast, wall-scaling normalization in Fig. \ref{Fig4}(\textit{a}) cannot account for differences observed between the ZPG, pipe, and channel cases even for the highest $Re_\tau$. The result indicates that the generalized $\ell_\epsilon$ in Eq. \eqref{Eq2_4} can be applied deep into the outer layer and beyond the limits of the logarithmic region where deviations from wall scaling begin to arise.

\begin{figure}
\centerline{\includegraphics{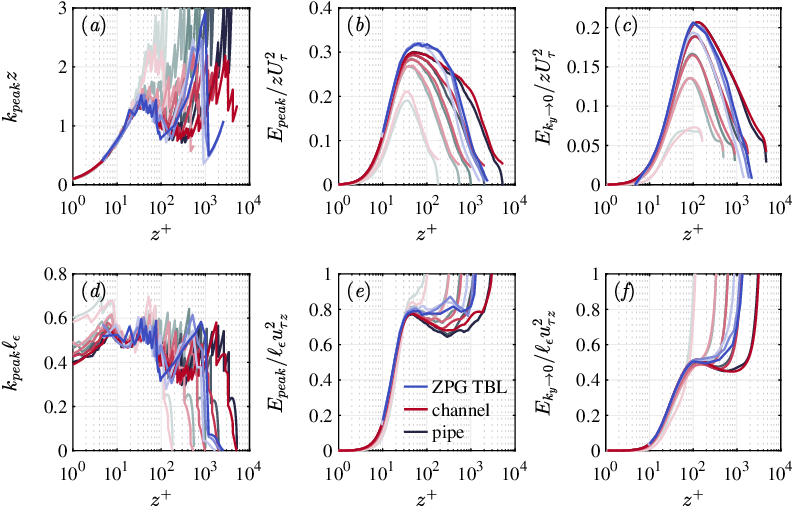}}
\caption{Profiles for the properties of $E_{ww}(k_y)$ identified in Fig. \ref{Fig2}(\textit{a}). Columns correspond to the wavenumber $k_{peak}$ of the spectrum peak (\textit{a,d}), the amplitude $E_{peak}$ of the peak (\textit{b,e}), and the low-wavenumber plateau $E_{k_y \to 0}$ (\textit{c,f}). Rows correspond to normalization with wall-scaling parameters (\textit{a,b,c}) and local-in-$z$ parameters (\textit{d,e,f}). The peak properties are detected directly from the spectra, and the low-wavenumber plateau is inferred from the fitted model von K{\'a}rm{\'a}n spectrum.}
\label{Fig5}
\end{figure}

Another noteworthy trend at $z/\delta=$ 0.5 is discrepancies at the lowest wavenumbers in Figure \ref{Fig4}(\textit{b}). The ZPG case contains significantly more low-wavenumber energy relative to $u_{\tau z}$ and $\ell_\epsilon$, such that there is no longer a true local peak at intermediate wavenumbers. The different low-wavenumber behavior in the ZPG TBL case is revisited in the later analysis.

Figure \ref{Fig5} shows wall-normal profiles of the $E_{ww}(k_y)$ spectrum properties discussed above, i.e. the wavenumber peak $k_{peak}$, energy density peak $E_{peak}$, and low-wavenumber limit $E_{k_y \to 0}$. The profiles for the ZPG TBL case are based on a limited number of $z$ positions compared to the channel and pipe cases. The properties $k_{peak}$ and $E_{peak}$ are estimated directly from the spectrum point with the highest energy density. The low-wavenumber limit is $E_{k_y \to 0}=a$, where $a$ is determined from fitting the Eq. \eqref{Eq2_1} von K{\'a}rm{\'a}n model to each spectrum as seen in Fig. \ref{Fig2}(\textit{b}).

The wavenumber peak $k_{peak}$ in Fig. \ref{Fig5}(\textit{a}) is proportional to the wall-normal distance $z$ for the highest $Re_\tau$ cases at intermediate distances beginning near $z^+\approx$ 30 corresponding to the start of $\ell_\epsilon \approx \kappa z$ in Fig. \ref{Fig3}(\textit{a}). In contrast, the peak $k_{peak}$ in Fig. \ref{Fig5}(\textit{d}) remains proportional to $\ell_\epsilon$ across a much wider range from the viscous sublayer to approximately 0.5$\delta$, and accounts for deviations from $z$ scaling in the lowest $Re_\tau$ cases. The wavenumber peak $k_{peak} \ell_\epsilon$ is not constant within this range, however, and varies moderately with wall-normal position including a weak decrease in the outer layer. This variability may be due to how lower and higher wavenumber regions of the spectrum can influence the peak. Specifically, the low-wavenumber energy grows significantly in the outer layer, where this growth appears to overlap with the peak in Fig. \ref{Fig4}(\textit{b}), causing the peak to shift to a lower wavenumber. In this context, $\ell_\epsilon$ is the predominant factor in determining $k_{peak}$, but is not the sole parameter.

The peak energy density $E_{peak}$ also follows the general parameters in Fig. \ref{Fig5}(\textit{e}) more closely than the wall scaling in Fig. \ref{Fig5}(\textit{b}). In particular, $\ell_\epsilon$ and $u_{\tau z}$ account for the Reynolds-number effects apparent in Figs. \ref{Fig4}(\textit{a}) and \ref{Fig5}(\textit{b}). The general parameters also account for the higher peak amplitude of the ZPG TBL case seen in Fig. \ref{Fig5}(\textit{b}). The initial increase of $E_{peak}$ within the near-wall viscous region in Fig. \ref{Fig5}(\textit{e}) is likely due in part to insufficient scale separation in the spectrum at these positions, where the dissipative cutoff will directly overlap with and dampen the peak energy at small $Re_\epsilon$, i.e. $f(k_{peak} \eta) < 1$ in Eq. \eqref{Eq2_1}. The normalized energy density decreases moderately in the outer layer before increasing again in the upper half of the boundary layer. The intermediate decrease is not fully understood at this time. The final noteworthy trend in Fig. \ref{Fig5}(\textit{e}) is that the ZPG TBL case has a greater peak in the outer layer, which is related to the low-wavenumber contribution and lower $k_{peak}$ discussed above.

The low-wavenumber limit $E_{k_y \to 0}$ in Fig. \ref{Fig5}(\textit{c,f}) follows the same trends as $E_{peak}$ in Fig. \ref{Fig5}(\textit{b,e}). The general parameters in Fig. \ref{Fig5}(\textit{e}) account for the differences in DNS cases within the near-wall region including the logarithmic layer. Consistent with the spectra in Fig.  \ref{Fig5}(\textit{b}) at $z/\delta=$ 0.5, the ZPG TBL flow has increased energy relative to $\ell_\epsilon$ and $u_{\tau z}$ in the outer layer. The largest $w$ motions are expected to scale with $\delta$ and $U_{\tau}$ based on studies of large-scale-motions in the literature \citep[see, e.g.][]{Guala2006,Balakumar2007}. However, this expectation is difficult to confirm with the present results because the majority of the $w$ spectrum is governed by the local scale $\ell_\epsilon \sim O(0.1\delta)$ which has limited separation from $\delta$ far from the wall (see, e.g. Fig. \ref{Fig3}(\textit{b)}), and $E_{k_y \to 0}$ additionally varies with $z/\delta$ in the outer layer regardless of the normalization.

In Fig. \ref{Fig5}(\textit{d,e,f}), there is an abrupt change in the spectral properties at positions far from the wall. The change occurs consistently at $z/\delta \approx$ 0.5 across all cases for each statistic (not shown here). Given that this position corresponds to the peak of $\ell_\epsilon$ in Fig. \ref{Fig3}(\textit{b}), it is assumed that $\ell_\epsilon$ loses its connection to the spectrum properties as its value decreases in the outer half of the boundary layer.

\subsection{Scale-decomposed variance of the wall-normal velocity}
\label{subsec:cond}

If the $w$ spectrum is proportional to $\ell_\epsilon$ and $u_{\tau z}$, these parameters will determine the overall variance corresponding to the integrated spectrum. Two additional factors will contribute to the variance: (1) the largest motions in the given spectrum, which vary in intensity between flow configurations as seen in Fig. \ref{Fig4}, and (2) the scale separation between the spectral peak and dissipative cutoff that determines the total small-scale energy. These two factors are evaluated here by computing the area under the spectrum both before and after the peak as classified in Fig. \ref{Fig2}(\textit{a}). The two integrals $\overline{{w^\prime}^2}_{l} = \int_0^{k_{peak}} E_{ww} dk_y$  and $\overline{{w^\prime}^2}_{s} = \int_{k_{peak}}^\infty E_{ww} dk_y$ sum to the total variance as $\overline{{w^\prime}^2} = \overline{{w^\prime}^2}_{l} + \overline{{w^\prime}^2}_{s}$. In this section, `large' and `small' refer to the size of turbulent motions relative to the spectrum wavenumber peak at a given wall-normal position.

The segregation of large and small scales using the spectral peak is considered a first-order estimate. This simplicity is preferred here due to the lack of tuning parameters. More sophisticated methods may be possible in the context of discussions presented in \S \ref{subsec:active}, but are left for future research. These methods would need to be applicable across the full boundary layer for a wide range of $Re_\tau$ to be of use for the present analysis. As an alternative, it may be possible to compute conditional statistics by integrating the model von K{\'a}rm{\'a}n spectrum defined in Eq. \eqref{Eq2_1} and shown in Fig. \ref{Fig2}. However, there is no clear analytical solution to the integral of this model when $f(k_y\eta)$ is included, and regardless the fitted model constants would need to be parameterized using the relevant physical scales.

\begin{figure}
\centerline{\includegraphics{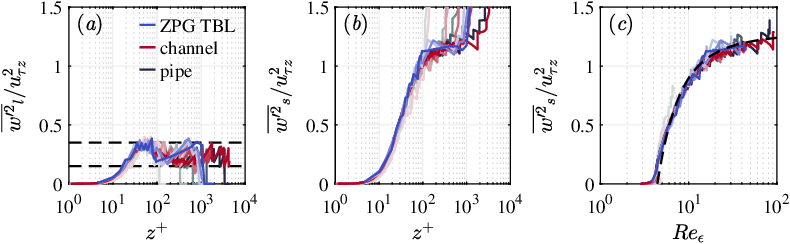}}
\caption{Contribution to the wall-normal variance from large-scale motions below the spectrum peak $\overline{{w^\prime}^2}_{l}$ and small-scale motions above the peak $\overline{{w^\prime}^2}_{s}$ as illustrated in Fig. \ref{Fig2}. (\textit{a}) wall-normal profile of $\overline{{w^\prime}^2}_{l}$, with dashed lines indicating the range 0.15 to 0.35. (\textit{b}) wall-normal profile of $\overline{{w^\prime}^2}_{s}$. (\textit{c}) $\overline{{w^\prime}^2}_{s}$ as a function of $Re_\epsilon$ for $z/\delta<$ 0.4, including the fitted function $1.29-1.5C_W Re_\epsilon^{-2/3}-18 Re_\epsilon^{-2}$ as a dashed black line.}
\label{Fig6}
\end{figure}

Wall-normal profiles of $\overline{{w^\prime}^2}_{l}$ and $\overline{{w^\prime}^2}_{s}$ -- based on the integrated spectrum at each $z$ position -- are shown in Fig. \ref{Fig6} relative to the local shear velocity $u_{\tau z}$. The vertical axis limits are matched in Fig. \ref{Fig6}(\textit{a,b}) to visualize the relative contribution of the two spectral regions. The contribution of large scales in Fig. \ref{Fig6}(\textit{a}) is relatively small and weakly variant throughout the boundary layer, where the observed range corresponding to the dashed lines is approximately bounded by $\overline{{w^\prime}^2}_{l}/u_{\tau z}^2 =$ 0.15 to 0.35. The small range and lack of prominent differences between flow configurations is attributed to two factors: limits in the lowest non-zero wavenumber which results from the domain size, and the method for estimating $\overline{{w^\prime}^2}_{l}$ which neglects low-wavenumber influences on the spectrum at and beyond the peak $k_y \ge k_{peak}$. No further parameterization or interpretation of Fig. \ref{Fig6}(\textit{a}) is given for these reasons.

The high-wavenumber variance contribution $\overline{{w^\prime}^2}_{s}$ generally aligns with a common curve in the lower half of the boundary layer in Fig. \ref{Fig6}(\textit{b}), and increases sharply in the upper half of the boundary layer where $\ell_\epsilon$ loses relevance. The increase with increasing $z$ is expected based on the greater scale separation and more extensive Kolmogorov region of the spectrum farther from the wall. Given the dependence of $k_{peak}$ on $\ell_\epsilon$ observed in Fig. \ref{Fig5}(\textit{d}), the Reynolds number $Re_\epsilon = \ell_\epsilon / \eta$ is a more direct representation than $z^+$ for the extent of the spectrum beyond the peak. Accordingly, $\overline{{w^\prime}^2}_{s}$ is plotted versus $Re_\epsilon$ in Fig. \ref{Fig6}(\textit{c}), where there is a modest improvement in the collapse of the DNS cases to a common curve. Similar to the limitations for $\overline{{w^\prime}^2}_{l}$, the estimate of $\overline{{w^\prime}^2}_{s}$ is imperfect due to the sharp division applied between large- and small-scale contributions.

The dependence on local parameters observed in Figs. \ref{Fig6}(\textit{a,b}) can be combined with theoretical forms of the spectrum to propose a semi-empirical expression for the wall-normal velocity variance. At a sufficiently large $Re_\epsilon$, an inertial subrange of scales will emerge in the spectrum with self-similar behavior following Kolomogrov's \citeyearpar{Kolmogorov1941} hypotheses:

\begin{equation}
E_{ww}(k_y) = C_w \epsilon^{2/3} k_y^{-5/3},
\label{Eq3_1}
\end{equation}

\noindent where $C_w \approx$ 0.65 is a constant \citep{Saddoughi1994}. The lower wavenumber limit of the inertial subrange has been shown to be well predicted by $\ell_\epsilon$ (see \S \ref{subsec:params} and references therein) such that $E_{ww} = C_w \epsilon^{2/3} \ell_\epsilon^{5/3}$ at $k_y=\ell_\epsilon^{-1}$. The observed alignment of the $E_{ww}$ peak with the general parameters in Fig. \ref{Fig5}, along with the definition in Eq. \eqref{Eq2_4}, leads to the same proportional relation:

\begin{equation}
E_{peak} \sim  u_{\tau z}^2 \ell_\epsilon \sim \epsilon^{2/3} \ell_\epsilon^{5/3}.
\label{Eq3_2}
\end{equation}

\noindent The definition of $\ell_\epsilon$ is therefore inherently consistent with the transition from the larger scales that determine $\epsilon$ to the inertial subrange scales that depend only on $\epsilon$ as seen in Eq. \eqref{Eq3_1}. The only external factor in correctly setting this transition is the velocity parameter used to define $\ell_\epsilon$ in Eq. \eqref{Eq2_4}, which must be chosen based on observed scaling in the spectrum.

In the high-Reynolds-number limit, the variance contribution $\overline{{w^\prime}^2}_{s}$ becomes governed by the inertial subrange behavior in Eq. \eqref{Eq3_1}:

\begin{equation}
\overline{{w^\prime}^2}_{s} \approx \int_{\ell_\epsilon^{-1}}^{\eta^{-1}} C_w \epsilon^{2/3} k_y^{-5/3} dk_y,
\label{Eq3_3}
\end{equation}

\noindent where the integral limits are the scales for the start and end of the inertial subrange. Equation \eqref{Eq3_3} neglects the shape of both the spectrum peak and dissipative cutoff which deviate from -5/3. Evaluating the integral and using the definition of $\ell_\epsilon$ in Eq. \eqref{Eq2_4} leads to

\begin{equation}
\overline{{w^\prime}^2}_{s} \approx u_{\tau z}^2 \left( A_3 - 1.5 C_w Re_\epsilon^{-2/3} \right),
\label{Eq3_4}
\end{equation}

\noindent where $A_3 = 1.5 C_w$ is relaxed to an undefined constant on account of the neglected peak shape. The start of the integral near $k_{peak} \approx 0.5 \ell_{\epsilon}$ also impacts the value of $A_3$. The variance within the Kolmogorov region of the spectrum becomes proportional to the local shear stress and $u_{\tau z}$ due to its appearance in the definition of $\ell_\epsilon$. In addition to setting the length where the -5/3 inertial subrange aligns with the peak scaling, $\ell_\epsilon$ also ensures the high-wavenumber variance contribution shares the same velocity scaling as the spectral peak.

Equation \eqref{Eq3_4} is fitted to the DNS cases in Fig. \ref{Fig6}(\textit{c}). The fit includes an additional empirical higher-order term $Re_\epsilon^{-2}$ to account for effects at moderate Reynolds numbers when the the shape of the dissipative cutoff previously neglected in Eq. \eqref{Eq3_3} has a substantial contribution to the variance. The choice of higher-order exponent is empirical owing to the lack of an analytical solution to the model spectrum and dissipative cutoff used here. Choosing a different (or multiple) higher-order term leads to similar goodness of fit and only small changes to $B_3$ that are within the uncertainty range of the low wavenumbers. The fit results in the high-wavenumber contribution $\overline{{w^\prime}^2}_{s} \approx u_{\tau z}^2 ( 1.3 - 1.5 C_w Re_\epsilon^{-2/3} - 18 Re_\epsilon^{-2})$, which aligns well with the DNS up to $z/\delta=$ 0.4 as seen in Fig. \ref{Fig6}(\textit{c}). Combining this fit with the low-wavenumber range $\overline{{w^\prime}^2}_{l}/u_{\tau z}^2 =$ 0.15 to 0.35 yields an approximation for the overall variance:

\begin{equation}
\frac{\overline{w^{\prime 2}}}{u_{\tau z}^2} = 1.55 \pm 0.1 - 1.5 C_w Re_\epsilon^{-2/3} - 18 Re_\epsilon^{-2}.
\label{Eq3_5}
\end{equation}

Equation \eqref{Eq3_5} -- informed by observed trends in the spectra -- indicates that the wall-normal variance should be computed from the local shear velocity $u_{\tau z}$ rather than the surface friction velocity $U_\tau$. This dependence is supported by the normalized spectra seen in Figs. \ref{Fig4} and \ref{Fig5}, and an explanation is given in \S \ref{subsec:active}.

\section{Discussion}
\label{sec:discuss}

Besides the observed dependence on $u_{\tau z}$ and $\ell_\epsilon$, Eq. \eqref{Eq3_5} also provides insight into the three main questions of the study given in \S \ref{sec:intro} and highlighted in Fig. \ref{Fig1}. These questions are discussed in the subsequent sections: the asymptotic limit of $B_3$ and its Reynolds-number dependence are further evaluated in \S \ref{subsec:reynolds}, the deviations observed for the ZPG-TBL flow are explained in \S \ref{subsec:gradient}, and the universality of $B_3$ for canonical wall-bounded flows is considered in \S \ref{subsec:active}.

\subsection{Reynolds-number dependence of wall-normal variances}
\label{subsec:reynolds}

In the context of the AEH prediction in Eq. \eqref{Eq1_1}, the new result in Eq. \eqref{Eq3_5} suggests a more generalized form for the variance can account for differences in Reynolds number and flow configuration:

\begin{equation}
\frac{\overline{w^{\prime 2}}}{u_{\tau z}^2} = B_3 - f_3(Re_\epsilon),
\label{Eq4_1}
\end{equation}

\noindent where the high-$Re$ limit is $B_3 \approx 1.55 \pm 0.1$ and the Reynolds-number dependence can be approximated as

\begin{equation}
f_3(Re_\epsilon) \approx 1.5 C_w Re_\epsilon^{-2/3} + 18 Re_\epsilon^{-2}.
\label{Eq4_2}
\end{equation}

\noindent The finite Reynolds-number dependence in Eq. \eqref{Eq4_2} should theoretically be the same for the streamwise and spanwise variances assuming the spectrum is isotropic at the Kolmogorov scales \citep{Spalart1988,Perry1990}, with the only difference being a lower constant $C_v$ for the spanwise $v$ component aligned with the spanwise wavenumber $k_y$ \citep{Saddoughi1994}.

\begin{figure}
\centerline{\includegraphics{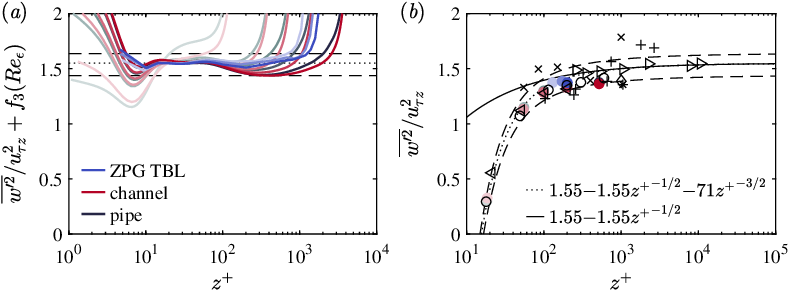}}
\caption{Evaluation of Eq. \eqref{Eq3_5} for the wall-normal variance. (\textit{a}) The DNS profiles compensated by the local stress $u_{\tau z}$ and the Reynolds number dependence $f_3(Re_\epsilon)$ in Eq. \eqref{Eq4_2}, with horizontal lines indicating $B_3=$ 1.55 (dotted) $\pm$0.1 (dashed). (\textit{b}) Variances measured at $z=0.1\delta$ for varying $Re_\tau$, where the lines are $B_3=$ 1.55 (dotted) $\pm$0.1 (dashed) with the correction $f_3(z^+)$ in Eq. \eqref{Eq4_3}, and the solid line excludes the higher order term. The markers in (\textit{b}) correspond to literature values: ($\bullet$) present DNS \citep{Sillero2013,Lee2015,Yao2023}; ($\circ$) pipe DNS \citep{Pirozzoli2021}; ($\Diamond$) channel DNS \citep{Hoyas2022}; ($\triangleleft$) Couette DNS \citep{Lee2018,Hoyas2024}; ($\triangleright$) pipe experiments \citep{Zhao2007}; ($+$) TBL experiments \citep{Fernholz1996}; ($\times$) TBL experiments \citep{Degraaff2000}; ($*$) TBL experiment \citep{Deshpande2020}.}
\label{Fig7}
\end{figure}

The revised variance prediction is evaluated for the DNS profiles in Fig. \ref{Fig7}(\textit{a}). The horizontal lines represent $B_3$ and the profiles are compensated by $f_3(Re_\epsilon)$ in Eq. \eqref{Eq4_2} to offset the variance ``missing'' due to the finite Reynolds number. The compensation leads to an observed collapse of the DNS along $B_3 \approx$ 1.55 in the range $z^+=$ 10 to 100, with discrepancies seen only for the cases with $Re_\tau=$ 180. Increases above the predicted value begin near $z/\delta \approx$ 0.5 for every case, consistent with the decrease in $\ell_\epsilon$ observed in Fig. \ref{Fig3}(\textit{b}) and the trends seen in Figs. \ref{Fig5} and \ref{Fig6}.

While Fig. \ref{Fig7}(\textit{a}) indicates promise for the revised $B_3$ and $f_3(Re_\epsilon)$, the profiles also exhibit modest variability at higher $z^+$ positions. Specifically, there is a greater decrease in the compensated variance above $z^+=$ 100 with increasing $Re_\tau$. The trend is consistent with the spectrum properties in Figs. \ref{Fig5}(\textit{a,b}). The shape of the curves in Fig. \ref{Fig7}(\textit{a}) are sensitive to at least two factors: (1) the changing contribution of large-scale energy across the boundary layer, and (2) the choice of the empirically fitted higher-order term in Eq. \eqref{Eq4_2}. A single empirical term is used here due to the modest range of $Re_\epsilon$ in Fig. \ref{Fig6}(\textit{c}) and the simplistic method employed for separating variance contributions which does not fully isolate the Reynolds-number dependence. Expanding Eq. \eqref{Eq4_2} to include additional terms would require a more sophisticated methodology and a correction that is informed by the shape of the dissipative cutoff. The results in Fig. \ref{Fig7}(\textit{a}) are therefore considered preliminary evidence in support of Eq. \eqref{Eq3_5}, with further tuning improvements left for future research.

As seen in Fig. \ref{Fig3}(\textit{c}), the Reynolds number can be approximated as $Re_\epsilon \approx (\kappa z^+)^{3/4}$ in the logarithmic layer. The Reynolds-number dependence in Eq. \eqref{Eq4_2} can therefore be expressed as a function of $z^+$ for cases with sufficiently high $Re_\tau$:

\begin{equation}
f_3(z^+) \approx 1.55 {z^+}^{-1/2} + 71 {z^+}^{-3/2}.
\label{Eq4_3}
\end{equation}

\noindent The same leading-order term $C{z^+}^{-1/2}$ has been used in previous spectrum-based estimates for the Reynolds-number dependence \citep{Perry1986}. However, the coefficient $C$ varies significantly across studies, including 0.65 \citep{Perry1986}, 1.55 (present), 4.37 \citep{Spalart1988}, and 5.58 \citep{Perry1990}.

The $f_3(z^+)$ relation in Eq. \eqref{Eq4_3} is combined with $B_3 \approx 1.5 5\pm 0.1$ in Fig. \ref{Fig7}(\textit{b}) to evaluate the present findings against previous measurements of wall-normal variances. The literature measurements defined in the Fig. \ref{Fig7} caption are specific to $z/\delta =$ 0.1 at the limit of the logarithmic region and are shown relative to $u_{\tau z}$. There is close agreement with $B_3-f_3(z^+)$ across the range of measurements, including for turbulent plane Couette flow which is not represented in the Table \ref{Tbl} cases. Most of the points are within the narrow uncertainty range corresponding to variability in large-scale energy seen in Fig. \ref{Fig6}(\textit{a}). The measurements outside of the dashed lines are all from laboratory experiments, which have a higher level of uncertainty due to challenges in measuring $w$ across all scales. A notable exception to the agreement in Fig. \ref{Fig7}(\textit{b}) is CICLoPE pipe measurements \citep{Orlu2017}, which are excluded here because the variances exceed 2 with local scaling and are considered to be an outlier.

To demonstrate the relative importance of the two terms in Eq. \eqref{Eq4_3}, the solid line in Fig. \ref{Fig7}(\textit{b}) represents $f_3(z^+)$ without the empirical $-3/2$ term. The higher-order correction representing the dissipative cutoff is significant for $Re_\tau \lesssim$ 1\,000 (i.e. $z^+ \lesssim$ 100 in the figure), but for $Re_\tau \approx$ 2\,800 its impact on the total wall-normal variance is less than 1\%, beyond which the $-1/2$ term representing the inertial subrange is dominant.

In addition to Eq. \eqref{Eq4_3} aligning with measurements in Fig. \ref{Fig7}(\textit{b}), the asymptotic limit $B_3 \approx$ 1.55 is within the literature range $B_3 = $ 1.5 to 1.85 discussed in \S \ref{sec:intro}. In particular, the value closely aligns with $B_3 \approx$ 1.55 fitted to measurements in the atmospheric surface layer \citep{Garratt1992}, indicating agreement in the DNS extrapolation to much larger Reynolds numbers. Atmospheric measurements are not included as individual data points in Fig. \ref{Fig7}(\textit{b}) due to significant scatter resulting from uncertainties including non-stationarity, thermal effects, and instrument limitations. While the agreement with atmospheric results is promising, more accurate measurements for $z^+ {\sim} O(10^5)$ are needed to improve confidence in the $B_3$ value.

\subsection{Wall-normal variances in enclosed and ZPG flows}
\label{subsec:gradient}

\begin{figure}
\centerline{\includegraphics{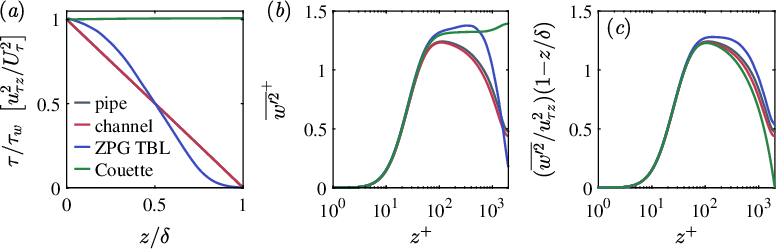}}
\caption{Profiles for flow cases at matched Reynolds number $Re_\tau \approx$ 2\,000 including DNS of a plane Couette flow \citep{Hoyas2024}. (\textit{a}) Stress $\tau = \rho u_{\tau z}^2$, where channel and pipe cases have the same linear decay. (\textit{b}) Variances relative to $U_\tau^2$. (\textit{c}) Variances relative to $u_{\tau z}^2/(1-z/\delta)$.}
\label{Fig8}
\end{figure}

The previous results and Eq. \eqref{Eq3_5} support the variance $\overline{w^{\prime 2}} \sim u_{\tau z}^2$ being proportional to the local stress rather than the surface scale $U_\tau^2$. This distinction is important when comparing statistics between different flow configurations that can have different stress profiles $\tau(z) / \tau_w = u_{\tau z}^2 / U_\tau^2$, with example profiles shown in Fig. \ref{Fig8}(\textit{a}). The profiles are at matched Reynolds number $Re_\tau \approx $ 2\,000 to minimize any differences due to Reynolds-number effects. Fig. \ref{Fig8} includes a turbulent plane Couette flow at the same Reynolds number \citep{Hoyas2024}. Couette flows are both enclosed and zero-pressure-gradient, thus allowing a direct assessment of factors that may lead to the different profile shapes for different canonical flows in Fig. \ref{Fig1}.

The channel and pipe flows in Fig. \ref{Fig8}(\textit{a}) both have linearly decaying stresses with $u_{\tau z}^2 = U_\tau^2 (1-z/\delta)$. In contrast, the ZPG-TBL flow has an approximately sigmoidal shape with $u_{\tau z} \approx U_{\tau}$ near the surface and the Couette flow has a nearly constant stress profile. These differing shapes result directly from the mean momentum balance at $z=0$, which reduces to $dP/dx = d \tau / dz$ with the stress decay balancing the mean pressure gradient. Equation \eqref{Eq1_2} and a constant near-surface stress is therefore specific to ZPG flows including Couette. In pressure-driven flows like a pipe or channel there is a non-negligible stress decay throughout the logarithmic region.

The different stress profiles can be compensated using $u_{\tau z}^2/(1-z/\delta)$ to normalize the variance \citep{Tuerke2013}. This normalization is identical to $U_\tau^2$ for a linear stress profile, and for the ZPG flows it compensates for the difference in $u_{\tau z}^2$ from the linear decay. The compensated variances in Fig. \ref{Fig8}(\textit{c}) show much closer agreement between all flow configurations than the surface normalization in Figs. \ref{Fig1}(\textit{a}) and \ref{Fig8}(\textit{b}). Compensating for the stress profile also accounts for the peak farther from the surface observed in Fig. \ref{Fig1}(\textit{b}). The higher $w$ variance and peak value farther from the wall can therefore be attributed predominately to the local stress dependence $\overline{w^{\prime 2}} \sim u_{\tau z}^2$ and the impact of the pressure gradient on the near-wall stresses.

The adjusted variance profile for the Couette flow in Fig. \ref{Fig8}(\textit{c}) matches closely with the other enclosed flows, but there are noteworthy differences in the spectrum. The spanwise spectrum for each velocity component exhibits a sharp peak for spanwise distances close to 5$\delta$ corresponding to domain-filling, counter-rotating rolls that are specific to Couette flows \citep{Komminaho1996,Lee2018,Hoyas2024}. See, e.g., Fig. 5 in \citet{Lee2018}. This peak significantly alters the scalewise energy distribution and dissipation such that the Couette spectra are not included here, yet the rolls appear to have minimal impact on the wall-normal variances in Fig. \ref{Fig8}(\textit{c}). It is speculated that these flow-specific structures lead to the same proportional increase in $\overline{w^{\prime 2}}$ and $\overline{u^\prime w^\prime}$ such that the wall-normal variance maintains the same relation to the local stress as other canonical flows.

The remaining differences between flows in Fig. \ref{Fig8}(\textit{c}) are small but non-negligible, most notably the higher ZPG-TBL peak. They cannot be attributed to effects of $f_3(Re_\epsilon)$ in Eq. \eqref{Eq4_2} because the $Re_\epsilon$ profiles in Fig. \ref{Fig3}(\textit{c}) overlap closely at the matched $Re_\tau$. The remaining differences are more likely due to the moderately larger footprint of turbulent eddies at low wavenumbers for the ZPG-TBL flow, which is evident from Figs. \ref{Fig4}(\textit{b}), \ref{Fig5}(\textit{f}), and \ref{Fig6}(\textit{a}). This footprint is discussed in further detail below.

\subsection{Active and inactive motions}
\label{subsec:active}

The analysis thus far utilized a primitive decomposition of large- and small-scale contributions to the variance shown in Fig. \ref{Fig2}(\textit{a}). For the purpose of this discussion, a more physically meaningful classification of the turbulent eddies is the concept of active and inactive motions \citep{Townsend1961}. Per Townsend's orginal hypothesis, at a given height $z_1$ in a population of superimposed attached eddies, active motions are centered near $z_1$ and contribute to the local Reynolds shear stress $\overline{u^\prime w^\prime}(z_1)$.  These motions, consequently, are expected to be directly proportional to the local stress. Inactive motions are larger and centered farther from the wall, and the portion of the eddy extending down to $z_1$ does not contribute to the correlation between $u$ and $w$ at $z_1$ \citep{Townsend1961,Deshpande2021}. 

Inactive motions contribute substantially to the streamwise and spanwise variance \citep{Deshpande2021}, but the AEH considers $w$ fluctuations to be purely active \citep{Townsend1961} with negligible inactive contributions \citep{Bradshaw1967,Katul1996}. The wall-normal $w^\prime$ fluctuations are thus closely related to the local shear stress, especially for the most energetic scales in the spectrum $E_{ww}$. It is therefore reasonable for these active motions to be proportional to $u_{\tau z}$ corresponding to the local stress. The dominance of active motions can then explain the observed dependencies for both the peak in the spectrum and the overall variance. This local stress is equivalent to the wall scaling in Eq. \eqref{Eq1_2} for ZPG flows in the high-$Re_\tau$ limit, but not for enclosed flows with a linear stress decay as seen in Fig. \ref{Fig8}(\textit{a}).

\begin{figure}
\centerline{\includegraphics{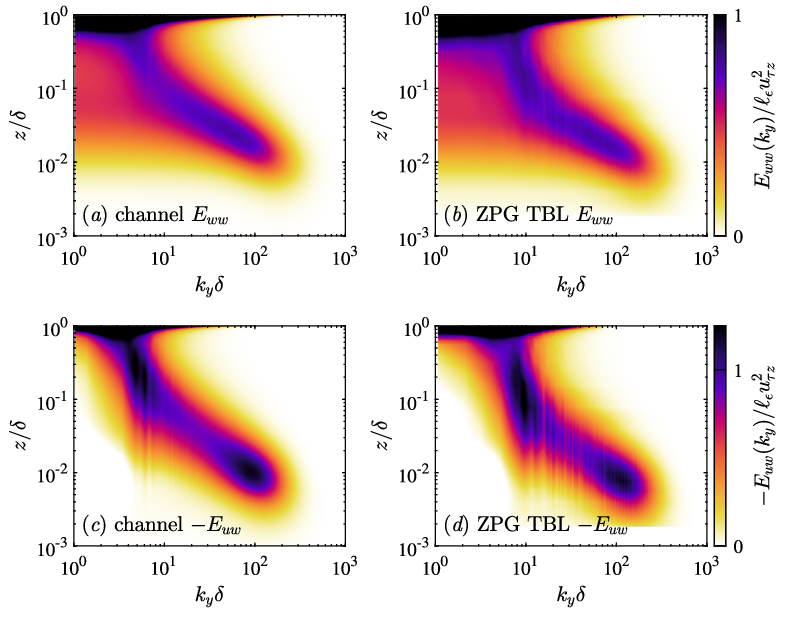}}
\caption{Spectra as a function of wavenumber $k_y$ and wall-normal position $z$ at matched Reynolds number $Re_\tau \approx$ 2000 for the channel and ZPG TBL cases. Rows correspond to $E_{ww}$ (top) and the shear stress spectrum $-E_{uw}$ (bottom).}
\label{Fig9}
\end{figure}

In its original form, Townsend's AEH predicts statistics for canonical wall-bounded flow based solely on a linear superposition of fluctuations, which emerge from a hierarchy of kinematic structures whose center scales with distance from the wall, i.e. geometrically self-similar eddies.
In real wall-bounded flows, however, these attached eddies coexist alongside non-self-similar motions including viscosity-dominated small scales and superstructures or very-large-scale motions (VLSMs), and there is non-linear interaction across all the motions \citep{Guala2011,Mckeon2017}. 
These non-linear interactions, primarily driven by `inactive' superstructures/VLSMs \citep{Deshpande2021}, lead to small $\delta$-scaled contributions to both the $w$ and $uw$ signals \citep{Deshpande2025}. 
This highlights a limitation of Townsend's original hypothesis, whereby inactive motions make a small yet non-negligible contribution to the wall-normal Reynolds-stress components that are considered purely active under Townsend's AEH. Further discussion on the interaction with inactive motions is given elsewhere \citep{Deshpande2025}.

These inactive contributions can be detected via an energy decomposition methodology \citep{Deshpande2021} requiring specific multi-point statistics that are not readily available for all the DNS datasets considered herein. The presence of active and inactive $w$ motions are instead inferred here by qualitatively comparing $E_{ww}$ spectra with the $E_{uw}$ shear stress spectra shown in Fig. \ref{Fig9} at matched Reynolds number for the channel and ZPG TBL cases. The peaks in the $E_{ww}$ and $E_{uw}$ spectra occur at similar wavenumbers and exhibit the same wall-normal trends in the lower half of the boundary layer. This similarity reaffirms the validity of \citeauthor{Townsend1976}'s (1976) hypothesis that the $w$ and $uw$ signals are dominated by contributions from the active (i.e. localized) motions.

While the scaling of the spectral peaks and surrounding wavenumbers is consistent between $E_{ww}$ and $E_{uw}$, the differences at the lowest wavenumbers for fixed $z/\delta$ positions in Fig. \ref{Fig4}(\textit{b}) show a clear failure of the local stress normalization at the largest scales. The contribution from the largest spanwise eddies to $E_{ww}(k_y)$ grows with distance from the wall as seen for wavenumbers $k_y \delta \lesssim$ 10 in Fig. \ref{Fig9}(\textit{a,b}), but there is an absence of energy at these same wavenumbers in Fig. \ref{Fig9}(\textit{c,d}) such that the largest eddies along $y$ do not necessarily contribute to the correlation of $u$ and $w$ at that same scale. Combined with the failure of the local stress scaling, this suggests that the low spanwise wavenumber contributions to $w$ and $uw$ are likely coming from the inactive motions. The contribution of these inactive motions to the overall variance, however, is small based on Fig. \ref{Fig6}(\textit{a}). On considering the energy distribution across frequency or streamwise wavenumber $k_x$ (not shown here), the spectra $E_{ww}(k_x)$ and $E_{uw}(k_x)$ are closely related even at the lowest wavenumber, suggesting the minor contribution of inactive motions is more difficult to discern from the dominant active motions in signals along time or $x$ \citep{Deshpande2021}. 

The weak signature of inactive motions in Fig. \ref{Fig4} corresponds with the wavenumbers of large- and very-large-scale motions, owing to which it is stronger in the outer region than in the lower log region. These motions are different for enclosed and non-enclosed flows \citep{Monty2009}, due to the influence of the opposite wall on global modes and the presence of a turbulent-nonturbulent interface in the latter case. The footprint of these motions varies with $z$, the type of flow and its Reynolds number, consistent with the trends seen in Fig. \ref{Fig5}(\textit{f}). It is therefore expected that the inactive motions, which deviate from the dependence on $u_{\tau z}$, are responsible for small discrepancies in $B_3$ including those seen in Fig. \ref{Fig8}(\textit{c}), thus precluding universality in the constant even at asymptotically high Reynolds numbers. Importantly, the inactive contributions become stronger at higher Reynolds numbers \citep{Deshpande2025}.

\subsection{An oversimplified model}

Based on the findings, it is instructive to simplify $E_{ww}(k_y)$ into 2 idealized regimes that replace Eq. \eqref{Eq2_1} for the purpose of identifying how the presence of inactive motions quantitatively changes the wall-normal variance. The low-wavenumber region imposes the leading-order behavior $E_{ww} \sim k_y^2$ in the vicinity of $k_y \to 0$, seen in the numerator of Eq. \eqref{Eq2_1}. The second region follows an idealized inertial subrange decay $C_w \epsilon^{2/3} k_y^{-5/3}$ up to $k_y \sim \eta^{-1}$, yielding a normalized composite spectrum

\begin{equation}
\frac{E_{ww}(k_y\ell_\epsilon)}{u_{\tau z}^2 \ell_\epsilon} =
    \begin{cases}
        A_{ia} + (C_w-A_{ia})k_y^2 & k_y < \ell_\epsilon^{-1}\\
        C_w k_y ^{-5/3} & \ell_\epsilon^{-1} \le k_y \le \eta^{-1}
    \end{cases}
    \label{Eq4_4}
\end{equation}

\begin{figure}
\centerline{\includegraphics{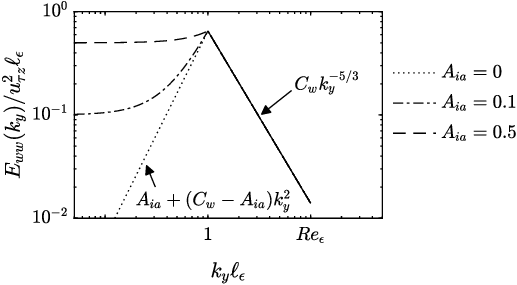}}
\caption{A simplified model spectrum with two idealized regions following Eq. \eqref{Eq4_4}, with examples for three $A_{ia}$ values representing the contribution of inactive motions.}
\label{Fig10}
\end{figure}

\noindent The transition point at $k_y \sim \ell_\epsilon^{-1}$ imposes a spectral peak at the same location as seen in Fig. \ref{Fig10}. The leading constant $a$ from Eq. \eqref{Eq2_1} is replaced here with $A_{ia}$ to represent the spectral amplitude of the large-scale and presumably inactive motions, and $(C_w - A_{ia})$ ensures the two regions match at the peak. When $A_{ia} = 0$ with no inactive contributions, the model in Fig. \ref{Fig10} resembles the stress spectra in Fig. \ref{Fig9}, and for finite $A_{ia}$ the shape is closer to the $w$ spectra in Fig. \ref{Fig4}. Integrating Eq. \eqref{Eq4_4} yields the idealized model variance

\begin{equation}
\frac{\overline{w^{\prime 2}}}{u_{\tau z}^2} = \frac{2}{3} A_{ia} + C_w \left[ \frac{11}{6} - \frac{3}{2} Re_\epsilon^{-2/3} \right], 
\label{eq:B3_ideal}
\end{equation}

\noindent where the high-Reynolds-number limit is $B_3 \approx 0.67 A_{ia} + 1.2$ following Eq. \eqref{Eq4_1} with $C_w =$ 0.65. The idealized variance can therefore range from $B_3 \approx$ 1.2 for $A_{ia}=$ 0 with no inactive contributions to $B_3 \approx$ 1.63 when $A_{ia}=C_w$ yields a low-wavenumber plateau as seen in Fig. \ref{Fig4} for the TBL flow at $z/\delta =$ 0.5.

This range matches closely with the $B_3$ uncertainty in the DNS datasets considering the simplicity of the idealized model. The oversimplified approximation of the variance in Eq. \eqref{eq:B3_ideal} and the discussed range demonstrate how a small inactive contribution $A_{ia}$ can lead to lower wall-normal variances even in the high-$Re_\tau$ limit. This reduction corresponding to lower large-scale inactive contributions was only speculated in the near-neutral atmospheric surface layer, where $B_3=$ 1.39 was reported \citep{Qin2025}.

\section{Summary}
\label{sec:summary}

In theory, Townsend's \citeyearpar{Townsend1976} Attached Eddy Hypothesis for high-Reynolds-number flows predicts a constant wall-normal velocity variance $B_3$ within the logarithmic region as seen in Eq. \eqref{Eq1_1}. In practice, $B_3$ depends on $Re_\tau$, varies weakly with wall-normal distance, and has a higher value for zero-pressure-gradient flows (Fig. \ref{Fig1}). These dependencies are investigated here using spectra $E_{ww}(k_y)$ as a function of spanwise wavenumber for DNS of channel, pipe, and ZPG flat plate boundary layers. This choice is motivated by the presence of a distinct peak at intermediate scales of the spanwise wavenumber spectra (Fig. \ref{Fig2}), which is not present in the streamwise wavenumber or frequency spectra.

The amplitude and wavenumber of the spectrum peak are proportional to the local shear velocity $u_{\tau z}$ (Eq. \ref{Eq2_3}) and dissipation-based length $\ell_\epsilon$ (Eq. \ref{Eq2_4}). This dependence persists from the viscous sublayer to approximately half the boundary layer depth (Figs. \ref{Fig4} and \ref{Fig5}). While these parameters are equivalent to the attached scales $U_\tau$ and $z$ in the near-wall region at high Reynolds numbers (Fig. \ref{Fig3}), the generalized length $\ell_\epsilon$ accounts for trends at small $Re_\tau$, is applicable across the entire lower half of the boundary layer, and transitions the spectrum directly into the inertial subrange which depends only on the local rate of dissipation (\S \ref{subsec:cond}).

The observed trends in the spectra result in the wall-normal variance depending on $u_{\tau z}$ and a local Reynolds number $Re_\epsilon=\ell_\epsilon/\eta$ (Fig. \ref{Fig6}\textit{c}). The dependence is approximated using a leading-order $Re_\epsilon^{-2/3}$ term derived from the inertial subrange and a higher-order empirical term to correct for the dissipative cutoff (Eq. \ref{Eq4_2}). The resulting semi-empirical expression aligns closely with the present DNS and other variance measurements in the literature (Fig. \ref{Fig7}), noting that further investigation is required to verify and refine the empirical fit. For predicting the variance in practical applications, the dependencies $u_{\tau z}$ and $Re_\epsilon$ in Eq. \eqref{Eq3_5} can be estimated as functions of $U_\tau$, $Re_\tau$, and $z/\delta$ that would differ between ZPG and pressure-driven flows.

Extrapolating Eq. \eqref{Eq3_5} to the limit $Re_\tau \to \infty$ suggests $B_3 \approx 1.55 \pm 0.1$, which is within the range of values previously reported and is closely aligned with empirical fits to atmospheric boundary layer measurements \citep{Panofsky1977}. The dependence of $B_3$ on $u_{\tau z}$ rather than $U_\tau$ alone is consistent with previous evidence \citep{Tuerke2013,Lozano2019} and is attributed to the wall-normal motions being predominately `active' per the terminology of \citeauthor{Townsend1976}'s (1976) attached eddy hypothesis, which ties $w$ to the local Reynolds shear stress. The hypothesis assumes a constant stress profile (Eq. 1.2), which is only achieved for ZPG flows in the high-$Re_\tau$ limit. The stress profile for each case, represented by $u_{\tau z}$, accounts for most of the differences in wall-normal variance observed across channel, pipe, ZPG-TBL, and plane Couette flows (Fig. \ref{Fig8}). Achieving convincing scaling for second-order statistics has been an ongoing challenge owing to dependencies on both local and global flow properties that can differ across canonical cases \citep{Marusic2010}, and the present work shows that the wall-normal component of the variance is predominately governed by the local flow properties.

The lowest wavenumbers in the $E_{ww}(k_y)$ spectrum contribute relatively little to the total variance (Fig. \ref{Fig6}\textit{a}), but this contribution increases with wall-normal distance and is greater for ZPG flows (Figs. \ref{Fig4}\textit{b} and \ref{Fig5}\textit{f}). The motions at these wavenumbers are speculated to be `inactive' owing to the absence of energy at the same wavenumbers in the shear stress spectrum (Fig. \ref{Fig9}) and the fact that the low-wavenumber energy is poorly described by both distance-from-the-wall and local flow parameters (Fig. \ref{Fig4}). The small yet non-negligible dependence of the variance on these inactive motions, which changes with distance and flow configuration, is a minor departure from Townsend's hypothesis that considers wall-normal fluctuations to be purely `active'. While the present results support the hypotheses for attached eddies and active motions to first-order accuracy, these dissimilar low-wavenumber contributions suggest there is a small range of possible $B_3$ values in the high-Reynolds-number limit rather than a single universal value applicable to all canonical wall-bounded flows.

\section*{Acknowledgments}
\begin{acknowledgments}
We gratefully acknowledge the authors of the original DNS studies for providing public access to the statistics and spectra that made this work possible: J. A. Sillero, M. Lee, J. Yao and their respective co-authors.
R. Deshpande is supported by the University of Melbourne's Postdoctoral Fellowship and acknowledges insightful discussions with Ivan Marusic.
\end{acknowledgments}

\section*{Declaration of interests}
The authors have no conflicts to disclose.

\section*{Data availability statement}
All data are available from the original DNS studies \citep{Sillero2013,Lee2015,Yao2023} and their corresponding datasets hosted online.

\bibliographystyle{jfm}
\bibliography{JFM-2025-2295_references}

\end{document}